\documentclass[ALICE,manyauthors]{cernphprep}

\usepackage[comma,square,numbers,sort&compress]{natbib}
\usepackage{hyperref}
\usepackage{lineno}
\usepackage{xcolor}
\usepackage{textcomp}
\usepackage[T1]{fontenc}
\usepackage{xspace}
\usepackage{orcidlink}

\newcommand{\rhopd}           {\ensuremath{\rho_{\mathrm{\overline{p}\overline{d}}}\xspace}}
\usepackage{xspace}

\newcommand{\PbPb}         {\mbox{Pb--Pb}\xspace}

\newcommand{\snn}          {\ensuremath{\sqrt{s_{\mathrm{NN}}}}\xspace}
\newcommand{\pt}           {\ensuremath{p_{\rm T}}\xspace}

\newcommand{\dEdx}         {\ensuremath{\textrm{d}E/\textrm{d}x}\xspace}

\newcommand{\dVdy}         {\ensuremath{\mathrm{d}V/\mathrm{d}y}\xspace}

\newcommand{\nineH}        {$\sqrt{s}~=~0.9$~Te\kern-.1emV\xspace}
\newcommand{\seven}        {$\sqrt{s}~=~7$~Te\kern-.1emV\xspace}
\newcommand{\twoH}         {$\sqrt{s}~=~0.2$~Te\kern-.1emV\xspace}
\newcommand{\twosevensix}  {$\sqrt{s}~=~2.76$~Te\kern-.1emV\xspace}
\newcommand{\five}         {$\sqrt{s}~=~5.02$~Te\kern-.1emV\xspace}
\newcommand{\twosevensixnn}{$\sqrt{s_{\mathrm{NN}}}~=~2.76$~Te\kern-.1emV\xspace}
\newcommand{\fivenn}       {$\sqrt{s_{\mathrm{NN}}}~=~5.02$~Te\kern-.1emV\xspace}

\newcommand{\GeVc}         {Ge\kern-.1emV/$c$\xspace}
\newcommand{\MeVc}         {Me\kern-.1emV/$c$\xspace}
\newcommand{\TeV}          {Te\kern-.1emV\xspace}
\newcommand{\GeV}          {Ge\kern-.1emV\xspace}
\newcommand{\MeV}          {Me\kern-.1emV\xspace}
\newcommand{\GeVmass}      {Ge\kern-.2emV/$c^2$\xspace}
\newcommand{\MeVmass}      {Me\kern-.2emV/$c^2$\xspace}

\newcommand{\pbar}         {\ensuremath{\rm\overline{p}}\xspace}

\newcommand{\dbar}   {\ensuremath{\overline{\mathrm{d}}}\xspace}
\newcommand{\dd}     {\ensuremath{\mathrm{d}}\xspace}

\newcommand{\dedx}{\mbox{\ensuremath{\dd E/\dd x}}\xspace}

\begin{document}%

\begin{titlepage}
\PHyear{2022}
\PHnumber{040}      
\PHdate{11 March}  
%

\title{First measurement of antideuteron number fluctuations at energies available at the Large Hadron Collider}
\ShortTitle{Antideuteron production fluctuations number at the Large Hadron Collider}   
\hyphenation{nu-cleo-syn-the-sis}
\Collaboration{ALICE Collaboration\thanks{See Appendix~\ref{app:collab} for the list of collaboration members}}
\ShortAuthor{ALICE Collaboration} 

\begin{abstract}
The first measurement of event-by-event antideuteron number fluctuations in high energy heavy-ion collisions is presented. The measurements are carried out at midrapidity ($|\eta| $ $<$ 0.8) as a function of collision centrality in \PbPb collisions at \snn = 5.02 TeV using the ALICE detector. A significant negative correlation between the produced antiprotons and antideuterons is observed in all collision centralities.
The results are compared with a state-of-the-art coalescence calculation. While it describes the ratio of higher order cumulants of the antideuteron multiplicity distribution, it fails to describe quantitatively the magnitude of the correlation between antiproton and antideuteron production. On the other hand, thermal-statistical model calculations describe all the measured observables within uncertainties only for correlation volumes that are different with respect to those describing proton yields and a similar measurement of net-proton number fluctuations.

\end{abstract}
\end{titlepage}
\setcounter{page}{2}

The production of nuclei and antinuclei in heavy-ion collisions has been extensively studied in the last two decades.
Nevertheless, this wealth of results is still not able to clarify the mechanism behind nuclei and antinuclei formation in heavy-ion collisions. Indeed, the two best fitting models, the coalescence~\cite{Mrowczynski:1987oid,Scheibl:1998tk,Sun:2018mqq} and the statistical hadronisation models (SHM)~\cite{Andronic:2010qu,Vovchenko:2018fiy}, give very similar predictions for the production rates of nuclei and antinuclei in heavy-ion collisions. This similarity calls for new observables to decisively discriminate between these two approaches.

The SHM describes the system as a hadron-resonance gas in thermal equilibrium at hadron emission, hence it predicts particle yields starting from the volume ($V$) and the temperature of the system at chemical freeze-out ($T_{\mathrm{chem}}$). The Grand Canonical Ensemble (GCE) formulation of the SHM fits the measured production yields of light hadrons and nuclei in central Pb--Pb collisions at center-of-mass energy ($\snn$) of 2.76 TeV with $T_{\mathrm{chem}}~=~$156.5 MeV~\cite{Andronic:2017pug}. The coalescence model uses a different approach to explain the production of nuclei: the size of the nucleon-emitting source, accessible through the analysis of femtoscopic correlations~\cite{ALICE:2020ibs}, the momentum distribution of the nucleons, as well as the nuclear wave function, are inputs that determine the formation probability of bound states~\cite{Blum:2017qnn,Sun:2018mqq}. While using statistical hadronisation it is possible to compute directly the absolute yields of particles, in the hadron coalescence model the yield of bound states can be computed only relative to the production of its components and as a function of system size.

In a recent model study~\cite{Feckova:2016kjx}, it is shown that the higher order cumulants of the deuteron yield distribution and correlation between proton (p) and deuteron (d) production can be used to distinguish between coalescence and SHM. Higher order cumulants $\kappa_{m}$ of the multiplicity distribution for $m$ $<$ 4 and the Pearson correlation coefficient ($\rho_{\mathrm{ab}}$) between different identified particles a and b can be expressed as
\begin{equation}
\label{eq7}
\kappa_{1} ~=~ \langle n \rangle,
\end{equation}
\begin{equation}
\label{eq8}
\kappa_{m} ~=~ \langle (n - \langle n \rangle)^{m} \rangle,
\end{equation}
\begin{equation}
\label{eq9}
\rho_{\mathrm{ab}} ~=~ \langle (n_{\mathrm{a}} - \langle n_{\mathrm{a}} \rangle)(n_{\mathrm{b}} - \langle n_{\mathrm{b}} \rangle) \rangle/\sqrt{\kappa_{2\mathrm{a}}\kappa_{2\mathrm{b}}},
\end{equation}
where $n$, $\langle n \rangle$, and $m$ are the event-by-event particle numbers, event average of particle numbers and order of the cumulants, respectively. The $\langle n_{\mathrm{a}} \rangle$ ($\langle n_{\mathrm{b}} \rangle$) and $\kappa_{2a}$ ($\kappa_{2b}$) are the first and second order cumulants of the multiplicity distribution of particle a (b).
In the GCE formulation of the SHM, the event-by-event deuteron multiplicity distribution is expected to follow the Poisson distribution~\cite{Braun-Munzinger:2020jbk}. Therefore various ratios between cumulants of different order of the deuteron multiplicity distribution such as $\kappa_{2}/\kappa_{1}$, $\kappa_{3}/\kappa_{2}$ are equal to unity in the GCE SHM. In a simple coalescence scenario, if deuterons are produced by the coalescence of thermally produced protons and neutrons, then the event-by-event deuteron distribution is expected to deviate from the Poisson baseline~\cite{Feckova:2016kjx}. By definition, the coalescence model also introduces a negative correlation between the measured proton and deuteron numbers in the absence of any initial correlation between proton and neutron. On the other hand, one does not expect any correlation between the measured p and d in the GCE SHM as the baryon productions from a thermal source are independent from each other. However, in the Canonical Ensemble (CE) formulation of the SHM, particle production is constrained by the conservation of the net baryon numbers on an event-by-event basis, which can also introduce a negative correlation between measured proton and deuteron in SHM and a deviation of cumulant ratios from the Poisson baseline~\cite{Barej:2020ymr, Braun-Munzinger:2020jbk}.

In this Letter, the first measurements of the $\kappa_{2}/\kappa_{1}$ ratio of antideuteron (antiparticles are used throughout the analysis to avoid the contamination from secondary deuterons coming from spallation processes in the beam pipe) multiplicity distribution and correlation ($\rho_{\mathrm{\pbar~\dbar}}$) between measured antideuterons (\dbar) and antiprotons (\pbar) are presented. Measurements are compared with predictions from the SHM and coalescence model in order to shed light on the deuteron synthesis mechanism. The results presented in this letter are obtained using data collected during the 2015 \PbPb LHC run at \snn$~=~$5.02 TeV.

The ALICE detector and its performance are described in detail in Refs.~\cite{Aamodt:2008zz,Abelev:2014ffa}. Collision events are selected by using the information from the V0C and V0A scintillator arrays~\cite{Abbas:2013taa}, located on both sides of the interaction point, covering the pseudorapidity intervals $-3.7 <\eta <-1.6$ and $2.8<\eta<5.1$, respectively. Events are selected with a minimum-bias (MB) trigger which requires at least one hit in both the V0A and the V0C detectors. In addition, only events with the primary vertex position within 10 cm along the beam axis to the nominal interaction point are selected to benefit from the full acceptance of the detector. Furthermore, to ensure the best possible performance of the detector and proper normalisation of the results, events with more than one reconstructed primary interaction vertex (pile-up events) are rejected. In total, about 100 million MB events are selected for analysis.
Furthermore, the selected events are divided into centrality classes based on the measured amplitude distribution in the V0A and V0C counters as described in Ref.~\cite{Abelev:2013qoq}. Central Pb--Pb collisions (head-on collisions) are obtained from the top 10$\%$ of the amplitude distribution corresponding to hadronic interactions and peripheral Pb--Pb collisions are obtained from the 70--80$\%$ region of the same distribution.

The charged-particle tracks are reconstructed in the ALICE central barrel with the Inner Tracking System (ITS)~\cite{Abelev:2014ffa} and the Time Projection Chamber (TPC)~\cite{Alme:2010ke}, which are located within a solenoid that provides a homogeneous magnetic field of up to $0.5$ T in the direction of the beam axis. These two subsystems provide full azimuthal coverage for charged-particle trajectories in the pseudorapidity interval $|\eta| < 0.8$. The transverse momentum range is restricted to 0.4 $<$ $p_{\mathrm{T}}$ $<$ 1.8 GeV/$c$ to select the \pbar and \dbar with high purity. Moreover, to guarantee a track-momentum resolution of $2\%$ in the relevant $p_{\mathrm{T}}$ range and an energy loss (\dedx) resolution in the TPC of $5\%$, the selected tracks are required to have at least 70 out of a maximum possible 159 reconstructed space points in the TPC, and at least one hit in the two innermost layers of the ITS. This selection also assures a resolution better than 300 $\mu$m~\cite{Abelev:2014ffa} on the distance of the closest approach to the primary vertex in the plane perpendicular (DCA$_{xy}$) and parallel (DCA$_z$) to the beam axis for the selected tracks. In addition, the $\chi^{2}$ per space point in the TPC and the ITS from the track fit are required to be less than 4 and 36, respectively. Daughter tracks from reconstructed secondary weak-decay kink topologies were rejected and a suppression of the weak-decay particles are obtained by selecting tracks with $|\mathrm{DCA}_z|$ and $|\mathrm{DCA}_{xy}|$ less than 1.0 and 0.1 cm, respectively.

The \dbar and \pbar are identified via the specific energy loss \dEdx in the gas volume of the TPC and the flight time of a particle from the primary vertex of the collision to the Time-of-Flight (TOF) detector. The $n(\sigma^{\rm TPC}_{i})$ variable represents the particle identification (PID) response in the TPC expressed in terms of the deviation between the measured and the expected \dEdx for a particle species $i$, normalized by the detector resolution $\sigma$. The expected \dEdx is computed with a parameterised Bethe--Bloch function~\cite{Abelev:2014ffa}. The \pbar and \dbar are identified using $-2~<~|n(\sigma^{\rm TPC}_{i})|~<$~4 in the range 0.4~$<~p_{\mathrm{T}}~<$~0.6~GeV/$c$ and 0.8~$<$~$p_{\mathrm{T}}$~$<$~1.0~GeV/$c$, respectively. Particle identification on a track-by-track basis using the TPC is limited to low momenta. Therefore, to identify \dbar (\pbar) in the range  1.0~$<$~$p_{\mathrm{T}}$~$<$~1.8~GeV/$c$ (0.6~$<$~$p_{\mathrm{T}}$~$<$~0.9~GeV/$c$), an additional selection of 3.0~$<$~$m^{2}$~$<$~4.2~GeV$^{2}$/$ c^{4}$ (0.6 $<$ $m^{2}$ $<$ 1.2 GeV$^{2}$/$c^{4}$) using the Time-of-Flight (TOF)~\cite{Akindinov:2013tea} detector is applied, where the square of the particle mass, $m^{2}$, is obtained by combining the information of the flight time with the trajectory length of the particle. The selection of \dbar is restricted to the range 0.8~$<$~$p_{\mathrm{T}}$~$<$~1.8 GeV/$c$ in order to keep the overall \dbar purity above 90$\%$. The \pbar selection is restricted to exactly half of the $p_{\mathrm{T}}$ range of \dbar according to the coalescence mechanism. This selection results in a purity of the selected \pbar sample above 95$\%$. The impurity in \dbar selection can lead to an autocorrelation with the selected \pbar and affect the \rhopd. The effect is negligible in our measurement as the \dbar and \pbar are mostly selected in separated \pt regions and in the common \pt interval the \dbar purity is $\sim99\%$. Selected \dbar and \pbar numbers in each event are further used to obtain the higher order cumulants and correlation. 

Measured cumulants are corrected for the \dbar and \pbar efficiencies assuming a binomial response of the detectors. The binomial-based method of efficiency correction~\cite{Nonaka:2017kko} is a two-step method. First, the efficiency of \dbar and \pbar reconstruction in the ALICE detector is obtained using a simulation based on GEANT4, which correctly describes the interaction of \pbar and \dbar with the material of the detectors~\cite{ALICE:2020zhb}. Then, the cumulants and correlation coefficient are corrected for the reconstruction efficiencies  using analytic expressions as discussed in Ref.~\cite{Nonaka:2017kko}. Typical reconstruction efficiencies of both \pbar and \dbar in the studied \pt ranges are about 70$\%$ and 25$\%$ in the TPC and TOF, respectively. The efficiency-corrected cumulants and correlation are further corrected for the centrality bin width effect~\cite{Luo:2013bmi} to suppress the initial volume fluctuations which arise from the initial state (size and shape) fluctuations.

The statistical uncertainties on the efficiency corrected $\kappa_{2}/\kappa_{1}$ ratio and $\rho_{\mathrm{\pbar~\dbar}}$ are obtained by the subsample method~\cite{ALICE:2017jsh}. The systematic uncertainties on the observables are estimated by varying the track selection and PID criteria. The systematic uncertainties due to track selection include the variation of the selection criteria on $\text{DCA}_{xy}$, $\text{DCA}_{z}$, the number of reconstructed space points in the TPC, and the quality of the track fit from their nominal values. The systematic uncertainties due to PID are calculated by varying the default $n(\sigma^{\rm{TPC}_{i}})$ and $m^{2}$ criteria. Systematic uncertainties due to each of these sources are considered as uncorrelated and the total systematic uncertainty on the observables is obtained by adding all the contributions in quadrature.

\begin{figure}
    \centering
    \includegraphics[width=0.6\textwidth]{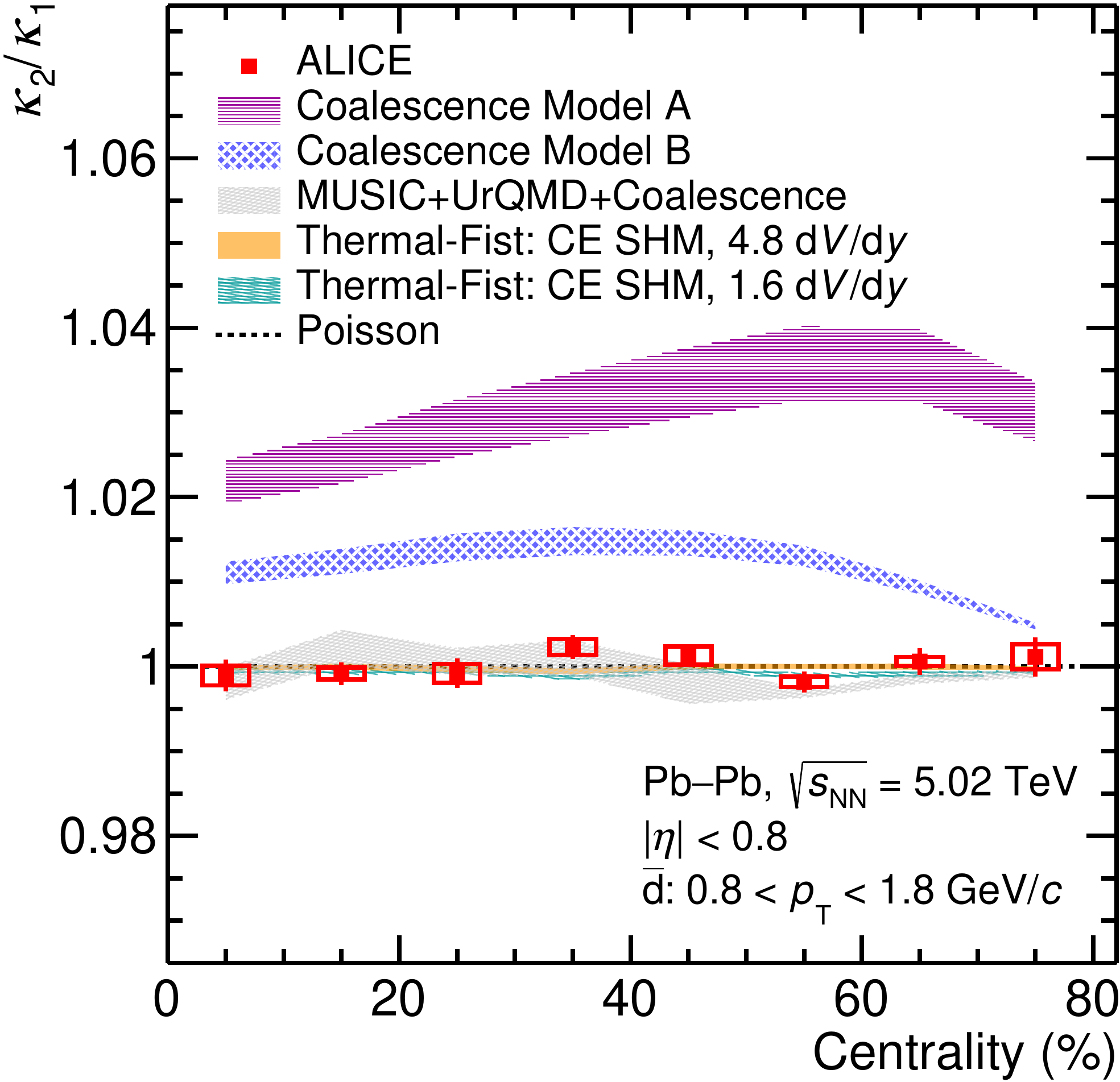}
    \caption{Second order to first order cumulant ratio of the \dbar multiplicity distribution as a function of collision centrality in Pb--Pb collisions at \snn$~=~$5.02 TeV. Statistical and systematic uncertainties are shown by the bars and boxes, respectively. Measured cumulant ratios are compared with estimations from the CE version of the SHM, from a simple coalescence model and from a MUSIC+UrQMD+Coalescence simulation. The width of the SHM model and MUSIC+UrQMD+Coalescence bands corresponds to the statistical uncertainty of the model estimation, whereas the width of the bands for the coalescence model corresponds to the uncertainty coming from the variation of the coalescence parameters.}
    \label{fig:C2C1}
\end{figure}

The resulting ratio of the second to first order cumulant for \dbar is shown in Fig.~\ref{fig:C2C1} for different centrality classes. The data is found to be consistent with unity within uncertainties as expected from a Poisson distribution and does not exhibit a significant centrality dependence. Measurements are also compared with estimations from the CE version of the SHM~\cite{Vovchenko:2019pjl} for two different correlation volumes ($V_\mathrm{c}$) for baryon number conservation, $V_\mathrm{c}~=~4.8$ \dVdy (orange band in figures) and $V_\mathrm{c}~=~1.6$ \dVdy (green band in figures). The choice of two different $V_\mathrm{c}$ is discussed below. In the SHM model the temperature is fixed to T$~=~155$ MeV~\cite{Vovchenko:2018fiy}, the volume fitted to the published pion, kaon, and proton yields at midrapidity~\cite{ALICE:2019hno}, and the net-baryon number set to 0. Measurements are found to be consistent with the SHM model for both of the $V_\mathrm{c}$. In contrast to the corresponding ratio for p and \pbar~\cite{STAR:2021iop, Vovchenko:2021kxx}, no strong dependence on the $V_\mathrm{c}$ is seen due to the fact that only a small fraction of the total antibaryon number is carried by \dbar~\cite{ALICE:2019nbs, Braun-Munzinger:2020jbk}. 
Remarkably, the data differs from the calculations of the coalescence model, which predicts a deviation larger than 1\% from the Poisson baseline as explained in Ref.~\cite{Feckova:2016kjx}. Two shaded bands are shown for the coalescence model: the purple one assumes full correlation among protons and neutrons produced in the collision (Model A), while the blue one assumes completely independent proton and neutron production fluctuations (Model B). On the other hand, a state of art model calculation coupling coalescence to a hydrodynamical model with hadronic interactions in the final state (MUSIC+UrQMD+COAL)~\cite{Sun:2022xsf} predicts $\kappa_{2}$/$\kappa_{1}$ ratio $\sim$ 1, in agreement with the experimental data (note that these predictions were updated after acceptance of this Letter). As discussed in~\cite{Sun:2022xsf}, the main difference between the coalescence predictions in Fig.~\ref{fig:C2C1} and the MUSIC+UrQMD+COAL calculation is due to the different method of implementing baryon number conservation.

\begin{figure}[h]
    \centering
    \includegraphics[width=0.6\textwidth]{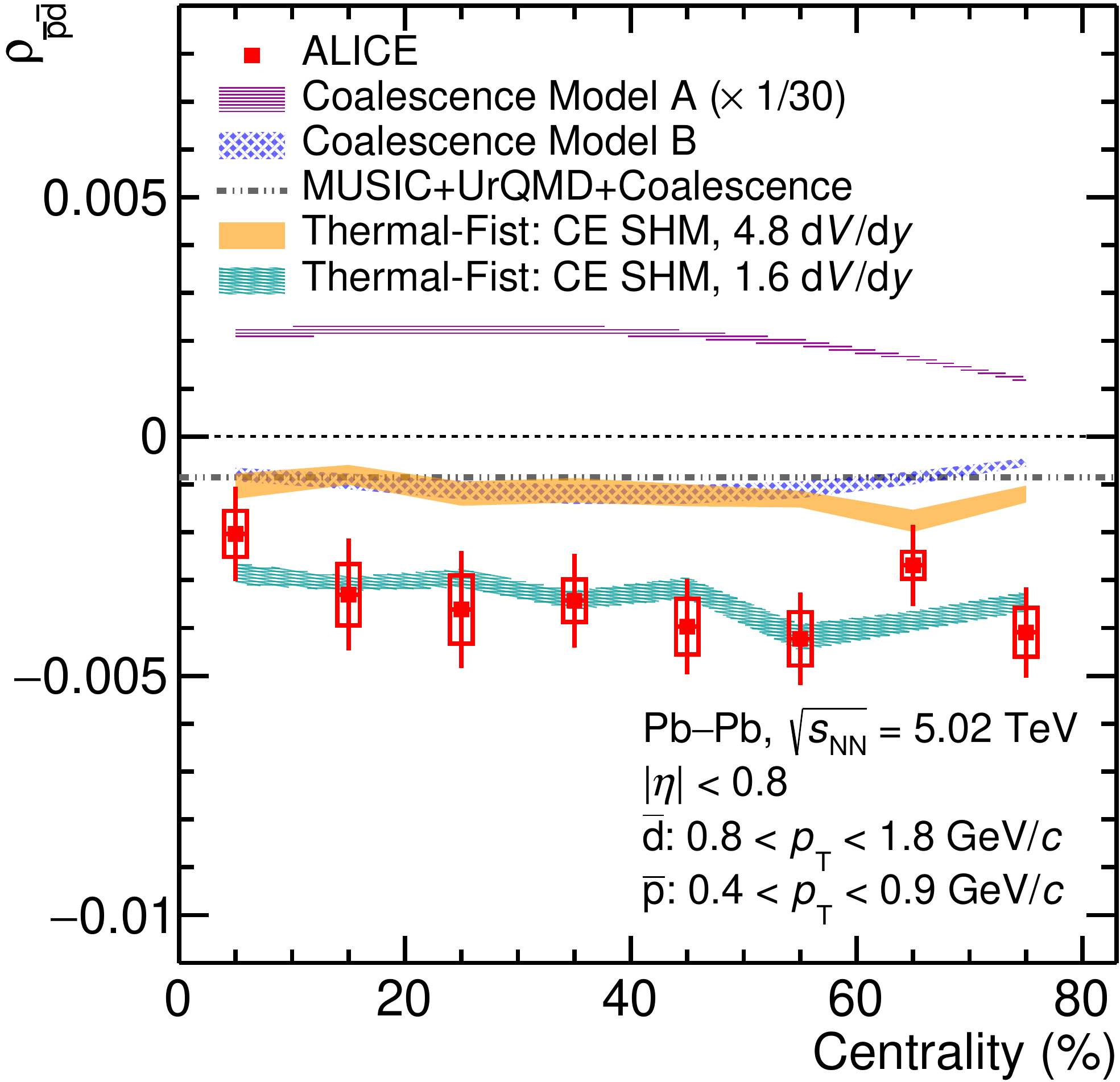}
    \caption{Pearson correlation between the measured \pbar and \dbar as a function of collision centrality in Pb--Pb collisions at \snn$~=~$5.02 TeV. Bars and boxes represent statistical and systematic uncertainties, respectively. Measured correlations are compared with estimations from the CE version of the SHM for two different baryon number conservation volumes, from coalescence model and from MUSIC+UrQMD+COAL.}
    \label{fig:rhovsCent}
\end{figure}

\begin{figure*}
    \centering
    \includegraphics[width=1.0\textwidth]{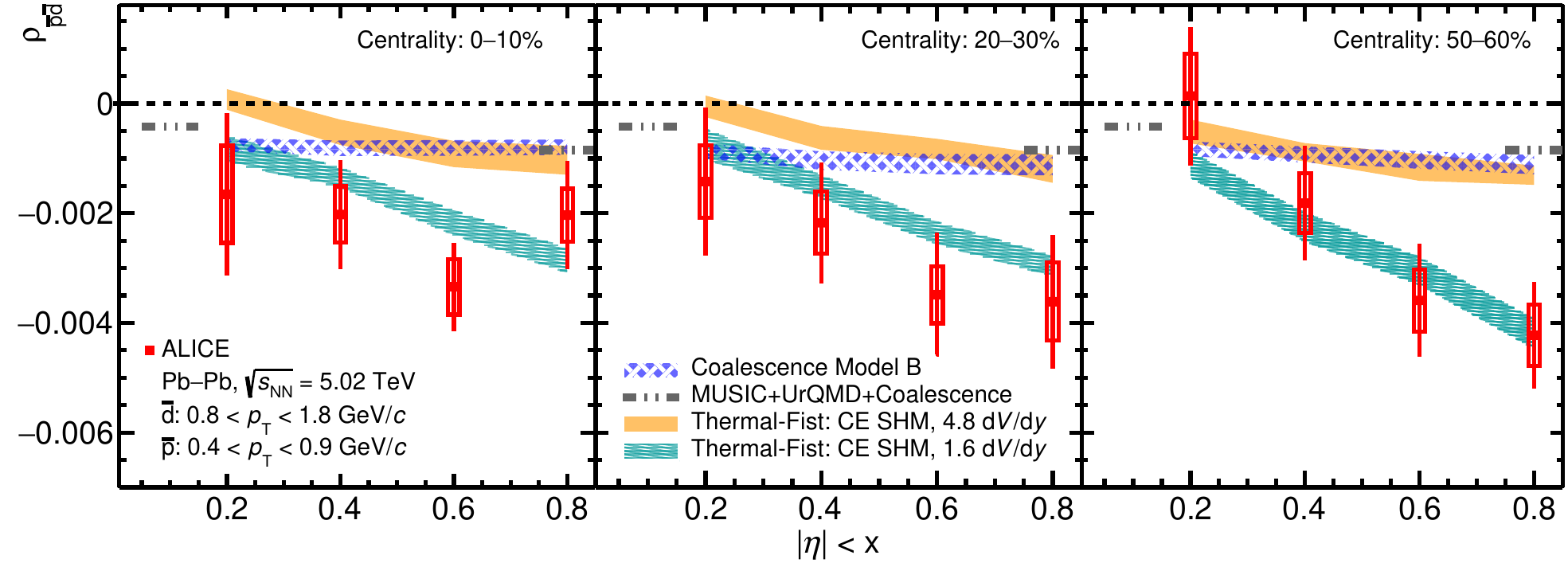}
    \caption{Dependence of \pbar--\dbar correlation on pseudorapidity acceptance of \pbar and \dbar selection in Pb--Pb collisions at \snn$~=~$5.02 TeV for three different centrality classes. Measurements are compared with calculations from the CE version of the SHM, coalescence model and MUSIC+UrQMD+COAL.}
    \label{fig:rhovseta}
\end{figure*}

Figure~\ref{fig:rhovsCent} shows \rhopd~ as a function of the collision centrality. A small negative correlation of $O$(0.1\%) is observed, i.e. in events with at least one \dbar, there are $O$(0.1\%) less \pbar observed than in an average event.
A negative correlation as observed in data is expected by the coalescence model (shown by the blue band in Fig.~\ref{fig:rhovsCent}) where \pbar and $\overline{\rm{n}}$ from two independent sources coalesce to produce \dbar. The same behaviour is observed for the MUSIC+UrQMD+COAL calculation.
It has to be noted that models based on fully correlated proton and neutron fluctuations (Model A in Ref.~\cite{Feckova:2016kjx}) predict values of $\rho$ around 6\% and are ruled out by data. On the other hand, the measured negative correlation between \pbar and \dbar is also expected by the CE version of the SHM which introduces a negative correlation between \pbar and \dbar through the conservation of a fixed net-baryon number. The predicted correlation in the SHM increases with decreasing correlation volume $V_\mathrm{c}$ for baryon number conservation which is used in the following for a determination of $V_\mathrm{c}$.
In order to determine the correlation volume for the baryon quantum number, a $\chi^2$ minimization is performed by varying the $V_\mathrm{c}$ parameter in the SHM model and comparing the result to the measured correlation as a function of centrality. 
The $V_\mathrm{c}$ interval probed in this case spans from 1 to 5 units of rapidity, and the value that describes best the measurement is $V_\mathrm{c}~=~1.6\pm 0.3$ \dVdy with a fit probability of 85\%. The SHM configuration with $V_\mathrm{c}~=~4.8$ \dVdy that correctly describes the net-proton number fluctuations in central \PbPb collisions~\cite{ALICE:2019nbs, Vovchenko:2020kwg} is compatible within  uncertainties with the measured \rhopd~only in central collisions. Conversely, this configuration is excluded with a 4$\sigma$ confidence level when compared with the measurements in all centrality classes.

Several consistency checks such as the correlation between \pbar and \dbar from different events, the correlation between antibaryon (\dbar) and baryon (p) were performed for a better understanding of the observed correlation. 
The correlation between \pbar and \dbar from mixed events is served as a null hypothesis test of the measurements and the obtained results are consistent with zero as expected.
However, a positive correlation is observed between antibaryon and baryon. This positive correlation is expected due to baryon number conservation~\cite{Braun-Munzinger:2020jbk}, whereas in simple coalescence model no correlation between baryon and antibaryon is expected as \dbar is not produced from the coalescence of p.

Figure~\ref{fig:rhovseta} shows the same Pearson correlation coefficient in three centrality intervals as a function of the $\eta$ acceptance of \pbar and \dbar selection. The observed anticorrelation is increasing with acceptance, and the effect is more pronounced for peripheral collisions. Simple coalescence calculations do not capture this trend. On the other hand, this measurement should motivate further calculations with more refined coalescence models. The decreasing trend seen in
the SHM with $V_\mathrm{c}~=~1.6$ \dVdy describes the experimental data. In the CE version of SHM model, anticorrelation between antibaryons depends on the fraction of antibaryon number in the acceptance out of the total conserved antibaryon numbers~\cite{Braun-Munzinger:2020jbk, Barej:2020ymr, Vovchenko:2020kwg, Vovchenko:2021kxx}. 
Therefore, the increased negative correlation magnitude with increasing acceptance can be understood as a consequence of baryon number conservation.

In summary, the measurement of \dbar production fluctuation is a valuable tool to challenge the nucleosynthesis models used for hadronic collisions. Simple coalescence models, as well as state-of-the-art MUSIC+UrQMD+COAL calculations, fail to fit simultaneously the measurement of the cumulant ratios and the correlation coefficient $\rho_{\mathrm{\pbar\dbar}}$. These models show a great sensitivity to the initial correlation between the proton and the neutron production, hence further theoretical developments might improve the comparison with the measurement. In recent studies, state-of-the-art CE SHM models are describing simultaneously proton yields and net-proton fluctuation measurements finding large $V_\mathrm{c}~\approx$ 3--5 \dVdy~\cite{Vovchenko:2020kwg, ALICE:2019nbs, Vovchenko:2019kes}. Surprisingly, deuteron production measurements~\cite{Vovchenko:2018fiy} as well as the fluctuation measurements presented here indicate a significantly smaller correlation volume for the baryon number. Under the assumption that $V_\mathrm{c}$ is independent of collision centrality, the value $V_\mathrm{c}~=~1.6\pm0.3$ \dVdy is obtained. This discrepancy might indicate a different production mechanism for light flavored hadrons and light nuclei. However, more sophisticated approaches including partial chemical equilibrium~\cite{Neidig:2021bal} or the implementation of the interaction of hadrons through phase shift~\cite{Andronic:2018qqt, Cleymans:2020fsc} could help in resolving this conundrum. The results of this Letter present a severe challenge to the current understanding of nuclei production in heavy-ion collisions at the LHC energies.

\newenvironment{acknowledgement}{\relax}{\relax}
\begin{acknowledgement}
\section*{Acknowledgements}

The ALICE Collaboration would like to thank all its engineers and technicians for their invaluable contributions to the construction of the experiment and the CERN accelerator teams for the outstanding performance of the LHC complex.
The ALICE Collaboration gratefully acknowledges the resources and support provided by all Grid centres and the Worldwide LHC Computing Grid (WLCG) collaboration.
The ALICE Collaboration acknowledges the following funding agencies for their support in building and running the ALICE detector:
A. I. Alikhanyan National Science Laboratory (Yerevan Physics Institute) Foundation (ANSL), State Committee of Science and World Federation of Scientists (WFS), Armenia;
Austrian Academy of Sciences, Austrian Science Fund (FWF): [M 2467-N36] and Nationalstiftung f\"{u}r Forschung, Technologie und Entwicklung, Austria;
Ministry of Communications and High Technologies, National Nuclear Research Center, Azerbaijan;
Conselho Nacional de Desenvolvimento Cient\'{\i}fico e Tecnol\'{o}gico (CNPq), Financiadora de Estudos e Projetos (Finep), Funda\c{c}\~{a}o de Amparo \`{a} Pesquisa do Estado de S\~{a}o Paulo (FAPESP) and Universidade Federal do Rio Grande do Sul (UFRGS), Brazil;
Ministry of Education of China (MOEC) , Ministry of Science \& Technology of China (MSTC) and National Natural Science Foundation of China (NSFC), China;
Ministry of Science and Education and Croatian Science Foundation, Croatia;
Centro de Aplicaciones Tecnol\'{o}gicas y Desarrollo Nuclear (CEADEN), Cubaenerg\'{\i}a, Cuba;
Ministry of Education, Youth and Sports of the Czech Republic, Czech Republic;
The Danish Council for Independent Research | Natural Sciences, the VILLUM FONDEN and Danish National Research Foundation (DNRF), Denmark;
Helsinki Institute of Physics (HIP), Finland;
Commissariat \`{a} l'Energie Atomique (CEA) and Institut National de Physique Nucl\'{e}aire et de Physique des Particules (IN2P3) and Centre National de la Recherche Scientifique (CNRS), France;
Bundesministerium f\"{u}r Bildung und Forschung (BMBF) and GSI Helmholtzzentrum f\"{u}r Schwerionenforschung GmbH, Germany;
General Secretariat for Research and Technology, Ministry of Education, Research and Religions, Greece;
National Research, Development and Innovation Office, Hungary;
Department of Atomic Energy Government of India (DAE), Department of Science and Technology, Government of India (DST), University Grants Commission, Government of India (UGC) and Council of Scientific and Industrial Research (CSIR), India;
National Research and Innovation Agency - BRIN, Indonesia;
Istituto Nazionale di Fisica Nucleare (INFN), Italy;
Japanese Ministry of Education, Culture, Sports, Science and Technology (MEXT) and Japan Society for the Promotion of Science (JSPS) KAKENHI, Japan;
Consejo Nacional de Ciencia (CONACYT) y Tecnolog\'{i}a, through Fondo de Cooperaci\'{o}n Internacional en Ciencia y Tecnolog\'{i}a (FONCICYT) and Direcci\'{o}n General de Asuntos del Personal Academico (DGAPA), Mexico;
Nederlandse Organisatie voor Wetenschappelijk Onderzoek (NWO), Netherlands;
The Research Council of Norway, Norway;
Commission on Science and Technology for Sustainable Development in the South (COMSATS), Pakistan;
Pontificia Universidad Cat\'{o}lica del Per\'{u}, Peru;
Ministry of Education and Science, National Science Centre and WUT ID-UB, Poland;
Korea Institute of Science and Technology Information and National Research Foundation of Korea (NRF), Republic of Korea;
Ministry of Education and Scientific Research, Institute of Atomic Physics, Ministry of Research and Innovation and Institute of Atomic Physics and University Politehnica of Bucharest, Romania;
Ministry of Education, Science, Research and Sport of the Slovak Republic, Slovakia;
National Research Foundation of South Africa, South Africa;
Swedish Research Council (VR) and Knut \& Alice Wallenberg Foundation (KAW), Sweden;
European Organization for Nuclear Research, Switzerland;
Suranaree University of Technology (SUT), National Science and Technology Development Agency (NSTDA), Thailand Science Research and Innovation (TSRI) and National Science, Research and Innovation Fund (NSRF), Thailand;
Turkish Energy, Nuclear and Mineral Research Agency (TENMAK), Turkey;
National Academy of  Sciences of Ukraine, Ukraine;
Science and Technology Facilities Council (STFC), United Kingdom;
National Science Foundation of the United States of America (NSF) and United States Department of Energy, Office of Nuclear Physics (DOE NP), United States of America.
In addition, individual groups or members have received support from:
Marie Sk\l{}odowska Curie, Strong 2020 - Horizon 2020 (grant nos. 824093, 896850), European Union;
Academy of Finland (Center of Excellence in Quark Matter) (grant nos. 346327, 346328), Finland;
Programa de Apoyos para la Superaci\'{o}n del Personal Acad\'{e}mico, UNAM, Mexico.
\end{acknowledgement}

\bibliographystyle{utphys}   
\bibliography{biblio}

\newpage
\appendix
\section{The ALICE Collaboration}
\label{app:collab}
\begin{flushleft} 
\small

S.~Acharya\,\orcidlink{0000-0002-9213-5329}\,$^{\rm 123,130}$, 
D.~Adamov\'{a}\,\orcidlink{0000-0002-0504-7428}\,$^{\rm 85}$, 
A.~Adler$^{\rm 68}$, 
G.~Aglieri Rinella\,\orcidlink{0000-0002-9611-3696}\,$^{\rm 32}$, 
M.~Agnello\,\orcidlink{0000-0002-0760-5075}\,$^{\rm 29}$, 
N.~Agrawal\,\orcidlink{0000-0003-0348-9836}\,$^{\rm 49}$, 
Z.~Ahammed\,\orcidlink{0000-0001-5241-7412}\,$^{\rm 130}$, 
S.~Ahmad\,\orcidlink{0000-0003-0497-5705}\,$^{\rm 15}$, 
S.U.~Ahn\,\orcidlink{0000-0001-8847-489X}\,$^{\rm 69}$, 
I.~Ahuja\,\orcidlink{0000-0002-4417-1392}\,$^{\rm 36}$
% ...

\section*{Affiliation Notes}

$^{\rm I}$ Deceased\\
$^{\rm II}$ Also at: Max-Planck-Institut f\"{u}r Physik, Munich, Germany\\
$^{\rm III}$ Also at: Italian National Agency for New Technologies, Energy and Sustainable Economic Development (ENEA), Bologna, Italy\\
$^{\rm IV}$ Also at: Dipartimento DET del Politecnico di Torino, Turin, Italy\\
$^{\rm V}$ Also at: Department of Applied Physics, Aligarh Muslim University, Aligarh, India\\
$^{\rm VI}$ Also at: Institute of Theoretical Physics, University of Wroclaw, Poland\\
$^{\rm VII}$ Also at: An institution covered by a cooperation agreement with CERN\\

\section*{Collaboration Institutes}

$^{1}$ A.I. Alikhanyan National Science Laboratory (Yerevan Physics Institute) Foundation, Yerevan, Armenia\\
$^{2}$ AGH University of Science and Technology, Cracow, Poland\\
$^{3}$ Bogolyubov Institute for Theoretical Physics, National Academy of Sciences of Ukraine, Kiev, Ukraine\\
$^{4}$ Bose Institute, Department of Physics  and Centre for Astroparticle Physics and Space Science (CAPSS), Kolkata, India\\
$^{5}$ California Polytechnic State University, San Luis Obispo, California, United States\\
$^{6}$ Central China Normal University, Wuhan, China\\
$^{7}$ Centro de Aplicaciones Tecnol\'{o}gicas y Desarrollo Nuclear (CEADEN), Havana, Cuba\\
$^{8}$ Centro de Investigaci\'{o}n y de Estudios Avanzados (CINVESTAV), Mexico City and M\'{e}rida, Mexico\\
$^{9}$ Chicago State University, Chicago, Illinois, United States\\
$^{10}$ China Institute of Atomic Energy, Beijing, China\\
$^{11}$ Chungbuk National University, Cheongju, Republic of Korea\\
$^{12}$ Comenius University Bratislava, Faculty of Mathematics, Physics and Informatics, Bratislava, Slovak Republic\\
$^{13}$ COMSATS University Islamabad, Islamabad, Pakistan\\
$^{14}$ Creighton University, Omaha, Nebraska, United States\\
$^{15}$ Department of Physics, Aligarh Muslim University, Aligarh, India\\
$^{16}$ Department of Physics, Pusan National University, Pusan, Republic of Korea\\
$^{17}$ Department of Physics, Sejong University, Seoul, Republic of Korea\\
$^{18}$ Department of Physics, University of California, Berkeley, California, United States\\
$^{19}$ Department of Physics, University of Oslo, Oslo, Norway\\
$^{20}$ Department of Physics and Technology, University of Bergen, Bergen, Norway\\
$^{21}$ Dipartimento di Fisica, Universit\`{a} di Pavia, Pavia, Italy\\
$^{22}$ Dipartimento di Fisica dell'Universit\`{a} and Sezione INFN, Cagliari, Italy\\
$^{23}$ Dipartimento di Fisica dell'Universit\`{a} and Sezione INFN, Trieste, Italy\\
$^{24}$ Dipartimento di Fisica dell'Universit\`{a} and Sezione INFN, Turin, Italy\\
$^{25}$ Dipartimento di Fisica e Astronomia dell'Universit\`{a} and Sezione INFN, Bologna, Italy\\
$^{26}$ Dipartimento di Fisica e Astronomia dell'Universit\`{a} and Sezione INFN, Catania, Italy\\
$^{27}$ Dipartimento di Fisica e Astronomia dell'Universit\`{a} and Sezione INFN, Padova, Italy\\
$^{28}$ Dipartimento di Fisica `E.R.~Caianiello' dell'Universit\`{a} and Gruppo Collegato INFN, Salerno, Italy\\
$^{29}$ Dipartimento DISAT del Politecnico and Sezione INFN, Turin, Italy\\
$^{30}$ Dipartimento di Scienze MIFT, Universit\`{a} di Messina, Messina, Italy\\
$^{31}$ Dipartimento Interateneo di Fisica `M.~Merlin' and Sezione INFN, Bari, Italy\\
$^{32}$ European Organization for Nuclear Research (CERN), Geneva, Switzerland\\
$^{33}$ Faculty of Electrical Engineering, Mechanical Engineering and Naval Architecture, University of Split, Split, Croatia\\
$^{34}$ Faculty of Engineering and Science, Western Norway University of Applied Sciences, Bergen, Norway\\
$^{35}$ Faculty of Nuclear Sciences and Physical Engineering, Czech Technical University in Prague, Prague, Czech Republic\\
$^{36}$ Faculty of Science, P.J.~\v{S}af\'{a}rik University, Ko\v{s}ice, Slovak Republic\\
$^{37}$ Frankfurt Institute for Advanced Studies, Johann Wolfgang Goethe-Universit\"{a}t Frankfurt, Frankfurt, Germany\\
$^{38}$ Fudan University, Shanghai, China\\
$^{39}$ Gangneung-Wonju National University, Gangneung, Republic of Korea\\
$^{40}$ Gauhati University, Department of Physics, Guwahati, India\\
$^{41}$ Helmholtz-Institut f\"{u}r Strahlen- und Kernphysik, Rheinische Friedrich-Wilhelms-Universit\"{a}t Bonn, Bonn, Germany\\
$^{42}$ Helsinki Institute of Physics (HIP), Helsinki, Finland\\
$^{43}$ High Energy Physics Group,  Universidad Aut\'{o}noma de Puebla, Puebla, Mexico\\
$^{44}$ Horia Hulubei National Institute of Physics and Nuclear Engineering, Bucharest, Romania\\
$^{45}$ Indian Institute of Technology Bombay (IIT), Mumbai, India\\
$^{46}$ Indian Institute of Technology Indore, Indore, India\\
$^{47}$ INFN, Laboratori Nazionali di Frascati, Frascati, Italy\\
$^{48}$ INFN, Sezione di Bari, Bari, Italy\\
$^{49}$ INFN, Sezione di Bologna, Bologna, Italy\\
$^{50}$ INFN, Sezione di Cagliari, Cagliari, Italy\\
$^{51}$ INFN, Sezione di Catania, Catania, Italy\\
$^{52}$ INFN, Sezione di Padova, Padova, Italy\\
$^{53}$ INFN, Sezione di Pavia, Pavia, Italy\\
$^{54}$ INFN, Sezione di Torino, Turin, Italy\\
$^{55}$ INFN, Sezione di Trieste, Trieste, Italy\\
$^{56}$ Inha University, Incheon, Republic of Korea\\
$^{57}$ Institute for Gravitational and Subatomic Physics (GRASP), Utrecht University/Nikhef, Utrecht, Netherlands\\
$^{58}$ Institute of Experimental Physics, Slovak Academy of Sciences, Ko\v{s}ice, Slovak Republic\\
$^{59}$ Institute of Physics, Homi Bhabha National Institute, Bhubaneswar, India\\
$^{60}$ Institute of Physics of the Czech Academy of Sciences, Prague, Czech Republic\\
$^{61}$ Institute of Space Science (ISS), Bucharest, Romania\\
$^{62}$ Institut f\"{u}r Kernphysik, Johann Wolfgang Goethe-Universit\"{a}t Frankfurt, Frankfurt, Germany\\
$^{63}$ Instituto de Ciencias Nucleares, Universidad Nacional Aut\'{o}noma de M\'{e}xico, Mexico City, Mexico\\
$^{64}$ Instituto de F\'{i}sica, Universidade Federal do Rio Grande do Sul (UFRGS), Porto Alegre, Brazil\\
$^{65}$ Instituto de F\'{\i}sica, Universidad Nacional Aut\'{o}noma de M\'{e}xico, Mexico City, Mexico\\
$^{66}$ iThemba LABS, National Research Foundation, Somerset West, South Africa\\
$^{67}$ Jeonbuk National University, Jeonju, Republic of Korea\\
$^{68}$ Johann-Wolfgang-Goethe Universit\"{a}t Frankfurt Institut f\"{u}r Informatik, Fachbereich Informatik und Mathematik, Frankfurt, Germany\\
$^{69}$ Korea Institute of Science and Technology Information, Daejeon, Republic of Korea\\
$^{70}$ KTO Karatay University, Konya, Turkey\\
$^{71}$ Laboratoire de Physique des 2 Infinis, Ir\`{e}ne Joliot-Curie, Orsay, France\\
$^{72}$ Laboratoire de Physique Subatomique et de Cosmologie, Universit\'{e} Grenoble-Alpes, CNRS-IN2P3, Grenoble, France\\
$^{73}$ Lawrence Berkeley National Laboratory, Berkeley, California, United States\\
$^{74}$ Lund University Department of Physics, Division of Particle Physics, Lund, Sweden\\
$^{75}$ Nagasaki Institute of Applied Science, Nagasaki, Japan\\
$^{76}$ Nara Women{'}s University (NWU), Nara, Japan\\
$^{77}$ National and Kapodistrian University of Athens, School of Science, Department of Physics , Athens, Greece\\
$^{78}$ National Centre for Nuclear Research, Warsaw, Poland\\
$^{79}$ National Institute of Science Education and Research, Homi Bhabha National Institute, Jatni, India\\
$^{80}$ National Nuclear Research Center, Baku, Azerbaijan\\
$^{81}$ National Research and Innovation Agency - BRIN, Jakarta, Indonesia\\
$^{82}$ Niels Bohr Institute, University of Copenhagen, Copenhagen, Denmark\\
$^{83}$ Nikhef, National institute for subatomic physics, Amsterdam, Netherlands\\
$^{84}$ Nuclear Physics Group, STFC Daresbury Laboratory, Daresbury, United Kingdom\\
$^{85}$ Nuclear Physics Institute of the Czech Academy of Sciences, Husinec-\v{R}e\v{z}, Czech Republic\\
$^{86}$ Oak Ridge National Laboratory, Oak Ridge, Tennessee, United States\\
$^{87}$ Ohio State University, Columbus, Ohio, United States\\
$^{88}$ Physics department, Faculty of science, University of Zagreb, Zagreb, Croatia\\
$^{89}$ Physics Department, Panjab University, Chandigarh, India\\
$^{90}$ Physics Department, University of Jammu, Jammu, India\\
$^{91}$ Physics Department, University of Rajasthan, Jaipur, India\\
$^{92}$ Physics Program and International Institute for Sustainability with Knotted Chiral Meta Matter (SKCM2), Hiroshima University, Hiroshima, Japan\\
$^{93}$ Physikalisches Institut, Eberhard-Karls-Universit\"{a}t T\"{u}bingen, T\"{u}bingen, Germany\\
$^{94}$ Physikalisches Institut, Ruprecht-Karls-Universit\"{a}t Heidelberg, Heidelberg, Germany\\
$^{95}$ Physik Department, Technische Universit\"{a}t M\"{u}nchen, Munich, Germany\\
$^{96}$ Politecnico di Bari and Sezione INFN, Bari, Italy\\
$^{97}$ Research Division and ExtreMe Matter Institute EMMI, GSI Helmholtzzentrum f\"ur Schwerionenforschung GmbH, Darmstadt, Germany\\
$^{98}$ Saha Institute of Nuclear Physics, Homi Bhabha National Institute, Kolkata, India\\
$^{99}$ School of Physics and Astronomy, University of Birmingham, Birmingham, United Kingdom\\
$^{100}$ Secci\'{o}n F\'{\i}sica, Departamento de Ciencias, Pontificia Universidad Cat\'{o}lica del Per\'{u}, Lima, Peru\\
$^{101}$ Stefan Meyer Institut f\"{u}r Subatomare Physik (SMI), Vienna, Austria\\
$^{102}$ SUBATECH, IMT Atlantique, Nantes Universit\'{e}, CNRS-IN2P3, Nantes, France\\
$^{103}$ Suranaree University of Technology, Nakhon Ratchasima, Thailand\\
$^{104}$ Technical University of Ko\v{s}ice, Ko\v{s}ice, Slovak Republic\\
$^{105}$ The Henryk Niewodniczanski Institute of Nuclear Physics, Polish Academy of Sciences, Cracow, Poland\\
$^{106}$ The University of Texas at Austin, Austin, Texas, United States\\
$^{107}$ Universidad Aut\'{o}noma de Sinaloa, Culiac\'{a}n, Mexico\\
$^{108}$ Universidade de S\~{a}o Paulo (USP), S\~{a}o Paulo, Brazil\\
$^{109}$ Universidade Estadual de Campinas (UNICAMP), Campinas, Brazil\\
$^{110}$ Universidade Federal do ABC, Santo Andre, Brazil\\
$^{111}$ University of Cape Town, Cape Town, South Africa\\
$^{112}$ University of Houston, Houston, Texas, United States\\
$^{113}$ University of Jyv\"{a}skyl\"{a}, Jyv\"{a}skyl\"{a}, Finland\\
$^{114}$ University of Kansas, Lawrence, Kansas, United States\\
$^{115}$ University of Liverpool, Liverpool, United Kingdom\\
$^{116}$ University of Science and Technology of China, Hefei, China\\
$^{117}$ University of South-Eastern Norway, Kongsberg, Norway\\
$^{118}$ University of Tennessee, Knoxville, Tennessee, United States\\
$^{119}$ University of the Witwatersrand, Johannesburg, South Africa\\
$^{120}$ University of Tokyo, Tokyo, Japan\\
$^{121}$ University of Tsukuba, Tsukuba, Japan\\
$^{122}$ University Politehnica of Bucharest, Bucharest, Romania\\
$^{123}$ Universit\'{e} Clermont Auvergne, CNRS/IN2P3, LPC, Clermont-Ferrand, France\\
$^{124}$ Universit\'{e} de Lyon, CNRS/IN2P3, Institut de Physique des 2 Infinis de Lyon, Lyon, France\\
$^{125}$ Universit\'{e} de Strasbourg, CNRS, IPHC UMR 7178, F-67000 Strasbourg, France, Strasbourg, France\\
$^{126}$ Universit\'{e} Paris-Saclay Centre d'Etudes de Saclay (CEA), IRFU, D\'{e}partment de Physique Nucl\'{e}aire (DPhN), Saclay, France\\
$^{127}$ Universit\`{a} degli Studi di Foggia, Foggia, Italy\\
$^{128}$ Universit\`{a} del Piemonte Orientale, Vercelli, Italy\\
$^{129}$ Universit\`{a} di Brescia, Brescia, Italy\\
$^{130}$ Variable Energy Cyclotron Centre, Homi Bhabha National Institute, Kolkata, India\\
$^{131}$ Warsaw University of Technology, Warsaw, Poland\\
$^{132}$ Wayne State University, Detroit, Michigan, United States\\
$^{133}$ Westf\"{a}lische Wilhelms-Universit\"{a}t M\"{u}nster, Institut f\"{u}r Kernphysik, M\"{u}nster, Germany\\
$^{134}$ Wigner Research Centre for Physics, Budapest, Hungary\\
$^{135}$ Yale University, New Haven, Connecticut, United States\\
$^{136}$ Yonsei University, Seoul, Republic of Korea\\
$^{137}$  Zentrum  f\"{u}r Technologie und Transfer (ZTT), Worms, Germany\\
$^{138}$ Affiliated with an institute covered by a cooperation agreement with CERN\\
$^{139}$ Affiliated with an international laboratory covered by a cooperation agreement with CERN.\\

\end{flushleft} 
\end{document}